\documentclass[%
reprint,
superscriptaddress,
 amsmath,amssymb,
 aps,
 pra,
]{revtex4-2}

\usepackage{graphicx}
\usepackage{dcolumn}
\usepackage{bm}
\usepackage{braket}
\usepackage{xcolor}
\usepackage{amsmath}
\usepackage{bbm}
\usepackage[colorlinks=true, citecolor=magenta, urlcolor=magenta, linkcolor=magenta]{hyperref}

\begin{document}

\title{Ground-State Sampling Correspondence\texorpdfstring{\\}{ }between Penalty-Based and Constrained Quantum Annealing}

\author{Shunta Ide}
\email{shunta16213@keio.jp}
\affiliation{Graduate School of Science and Technology, Keio University, 3-14-1, Hiyoshi, Kohoku-ku, Yokohama, Kanagawa 223-8522, Japan}
\author{Shu Tanaka}
\email{shu.tanaka@keio.jp}
\affiliation{Graduate School of Science and Technology, Keio University, 3-14-1, Hiyoshi, Kohoku-ku, Yokohama, Kanagawa 223-8522, Japan}
\affiliation{Department of Applied Physics and Physico-Informatics, Keio University, 3-14-1, Hiyoshi, Kohoku-ku, Yokohama, Kanagawa 223-8522, Japan}
\affiliation{Keio University Sustainable Quantum Artificial Intelligence Center (KSQAIC), Keio University,\\ 2-15-45, Mita, Minato-ku, Tokyo 108-8345, Japan}
\affiliation{Human Biology-Microbiome-Quantum Research Center (WPI-Bio2Q), Keio University,\\ 2-15-45, Mita, Minato-ku, Tokyo 108-8345, Japan}
\affiliation{Green Computing Systems Research Organization, Waseda University, 27, Waseda-cho, Shinjuku-ku, Tokyo 162-0042, Japan}

\date{\today}

\begin{abstract}
  Quantum annealing generally samples degenerate ground states with nonuniform probabilities.
  This unfair sampling depends on how the driver Hamiltonian couples these states near the end of annealing.
  For combinatorial optimization problems with a cardinality constraint, the penalty method and constrained quantum annealing (CQA) differ in these couplings.
  The transverse-field driver of the penalty method does not directly couple feasible ground states, whereas the fully connected XY driver of CQA couples pairs of them related by a single spin exchange.
  We investigate how this difference affects ground-state sampling, using closed-system simulations of random graph bipartitioning instances together with degenerate perturbation theory.
  For the instances studied, the conditional ground-state distribution of penalty-based quantum annealing approaches that of CQA at sufficiently large penalties and long annealing times.
  This convergence can be explained by the effective Hamiltonian of penalty-based quantum annealing, in which virtual single-spin flips to infeasible states generate a second-order interaction proportional to the CQA driver projected onto the feasible ground-state manifold, which is the leading term of the effective Hamiltonian of CQA.
  In the numerical simulations, the Jensen--Shannon divergence between the conditional ground-state distributions of penalty-based quantum annealing and CQA decreases approximately as the inverse square of the penalty-to-objective energy ratio, consistent with the perturbative analysis.
  This correspondence between the effective Hamiltonians extends to equal-weight fixed-magnetization constraints under the same perturbative conditions, providing a common mechanism for sampling bias in the two implementations.
\end{abstract}

\maketitle

\section{Introduction}
Quantum annealing (QA) is a heuristic for finding the ground state of an Ising model~\cite{finnila1994quantum,kadowaki1998quantum,farhi2001quantum,santoro2002theory}.
Because many combinatorial optimization problems (COPs) can be formulated as finding the ground state of an Ising model~\cite{lucas2014ising}, QA has been explored in a wide range of fields, including materials design~\cite{kitai2020designing,endo2022phase,inoue2022towards,xu2025quantum}, finance~\cite{tanahashi2019application,orus2019forecasting,takahashi2026effectiveness}, and logistics~\cite{inoue2021traffic,bao2023ising}.
In most applications, it is sufficient to find any one optimal solution.
Some applications, however, require all optimal solutions or as many distinct ones as possible~\cite{azinovic2017assessment,orus2019forecasting,sahai2020estimating}.
This task is considerably more demanding than finding a single solution, since even counting the solutions is \#P-complete for many NP-complete problems~\cite{valiant1979complexity}.
For such purposes, the performance of QA should be evaluated not only by the probability of obtaining a ground state but also by the probability distribution over the degenerate ground states.
Ideally, every ground state should be obtained with equal probability.

Even when the annealing time is long enough for the system to reach the ground-state manifold, QA generally samples the degenerate ground states with nonuniform probabilities, a phenomenon known as unfair sampling~\cite{matsuda2009ground,king2016degeneracy,mandra2017exponentially,zhang2017advantages,konz2019uncertain,maruyama2026graph,maruyama2026interpretation,maruyama2026uniformity,ide2026unfair}.
In the worst case, the sampling probabilities of some ground states are exponentially suppressed, and those states are therefore practically never obtained~\cite{matsuda2009ground,mandra2017exponentially}.
This unfair sampling has been analyzed using degenerate perturbation theory~\cite{boixo2013experimental,konz2019uncertain,maruyama2026interpretation}.
In this picture, the residual driver term near the end of annealing is treated as a perturbation to the problem Hamiltonian within its degenerate ground-state manifold~\cite{amin2009first}.
When the evolution is sufficiently slow to follow the instantaneous ground state within the dynamically accessible subspace, the lowest eigenstate of the resulting effective Hamiltonian determines which ground states are favored, provided any residual degeneracy is resolved.
This selection by quantum fluctuations is analogous to quantum order by disorder in frustrated spin systems~\cite{moessner2000two,tanaka2000quantum,moessner2001ising}.
The sampling bias therefore depends on both the driver and the energy landscape of the problem Hamiltonian.

Most practical COPs involve constraints.
In QA, such constraints must be incorporated into either the problem Hamiltonian or the driver Hamiltonian.
A common example is the cardinality constraint, which fixes the number of up spins and hence the total magnetization along the \( z \) axis.
For this constraint, there are two representative methods, the penalty method~\cite{lucas2014ising} and constrained quantum annealing (CQA)~\cite{hen2016quantum,hen2016driver}.
The penalty method keeps the conventional transverse-field driver and adds to the problem Hamiltonian a term that penalizes constraint violations~\cite{lucas2014ising}.
It is compatible with existing hardware, although the penalty coefficient must be chosen carefully~\cite{lucas2014ising,yin2024penalty,ide2025extending}.
CQA instead employs a driver that commutes with the constraint operator, such as the XY driver for cardinality constraints, and preserves feasibility when the initial state lies in the feasible subspace~\cite{hen2016quantum,kudo2018constrained}.
The two methods thus differ in the structure of both the driver and the problem Hamiltonian, and hence in the effective Hamiltonian that determines the sampling bias.

How this difference affects the conditional ground-state distributions produced by the two methods has not been clarified.
Previous work on weighted graph bipartitioning has shown that the penalty coefficient affects sampling bias and that increasing it can improve fairness~\cite{ide2026unfair}, but it did not establish how the resulting distribution is related to that of CQA.
When the penalty is large enough that the penalized problem Hamiltonian has no infeasible ground states, both methods target the same optimal feasible configurations.
This raises the question of under what conditions the two methods produce similar conditional ground-state distributions despite their different transition mechanisms.

In this paper, we compare penalty-based QA with a transverse-field driver and CQA with a fully connected XY driver for the graph bipartitioning problem (GBP), a representative COP with a cardinality constraint.
We combine degenerate perturbation theory with closed-system simulations of finite ensembles of random GBP instances.
In these simulations, the conditional ground-state distribution of penalty-based QA approaches that of CQA on average at large penalties and long annealing times.
This agreement does not imply uniform sampling.

Degenerate perturbation theory shows that, near the end of annealing, virtual single-spin flips through infeasible states generate the leading QA interaction at second order in the transverse field.
At large penalties, this interaction is proportional to the XY driver projected onto the feasible ground-state manifold.
The same projected operator appears at first order in CQA.
Consequently, when this projected operator has a unique lowest eigenstate within the dynamically accessible subspace, the two methods favor the same ground states.
The same operator correspondence holds for equal-weight constraints that fix the total magnetization to a given value.

\section{Model and Methods}\label{sec:model}
We compare two implementations of QA for the same constrained optimization problem.
The first uses a transverse-field driver and a quadratic penalty~\cite{lucas2014ising}, whereas the second uses a fully connected XY driver and is known as CQA~\cite{hen2016quantum}.
In the comparisons below, QA without qualification denotes the penalty-based method.
For each system size, we simulate both methods on the same set of random GBP instances over a range of annealing times, and we vary the penalty coefficient for penalty-based QA.
From the simulated final states, we evaluate the ground-state probability, the uniformity of the conditional distribution over the degenerate ground-state manifold, and the agreement between the conditional ground-state distributions produced by the two methods.
We also use degenerate perturbation theory to derive the effective Hamiltonians of the two methods near the end of annealing and compare them.
Based on this analysis, we examine how the agreement between the simulated distributions depends on the ratio of the penalty coefficient to the objective-energy scale.

\subsection{Graph bipartitioning and constraint implementation}
Consider an undirected weighted graph \( G=(\mathcal{V},\mathcal{E}) \) with vertex set \( \mathcal{V}=\{1,\ldots,N\} \), where \( N \) is even, and edge set \( \mathcal{E} \).
Each edge \( \{i,j\}\in\mathcal{E} \) carries a positive weight \( w_{ij}>0 \).
The graph bipartitioning problem (GBP) seeks a partition of \( \mathcal{V} \) into two equally sized subsets that minimizes the total weight of the edges between them.
GBP is known to be NP-hard~\cite{garey1974some,sotirov2018graph}.
We choose GBP because all instances at a fixed \( N \) share the same balance constraint and feasible configuration space, while the graph determines the objective function.
This allows us to compare the two constraint implementations across different objective landscapes while keeping the constraint fixed.

We assign a classical spin \( s_i\in\{-1,+1\} \) to each vertex \( i \) to indicate its group membership and denote a spin configuration by \( \bm{s}=(s_1,\ldots,s_N) \).
Let \( \sigma_i^x \), \( \sigma_i^y \), and \( \sigma_i^z \) denote the Pauli operators acting on the \( i \)th qubit, \( I \) the identity operator, and \( \ket{\bm{s}} \) the computational-basis state satisfying \( \sigma_i^z\ket{\bm{s}}=s_i\ket{\bm{s}} \) for all \( i \).
The objective Hamiltonian and balance constraint are
\begin{align}
  H_{\mathrm{obj}} & = \sum_{\{i,j\}\in\mathcal{E}}\frac{w_{ij}}{2}\left(I-\sigma_i^z\sigma_j^z\right),\label{eq:objective} \\
  \sum_{i=1}^N s_i & = 0.
\end{align}
The operator \( H_{\mathrm{obj}} \) is diagonal in the computational basis, and each term in Eq.~\eqref{eq:objective} gives \( w_{ij} \) when vertices \( i \) and \( j \) belong to different groups and zero otherwise.
Each eigenvalue of \( H_{\mathrm{obj}} \) is therefore the total weight of the edges between the two groups and is nonnegative.
Let \( S_z=\sum_{i=1}^N \sigma_i^z \) denote the total magnetization operator along the \( z \) axis.
The set of feasible configurations satisfying the balance constraint is
\begin{equation}
  \mathcal{F}=\left\{\bm{s}\in\{-1,+1\}^N \mid \sum_{i=1}^N s_i=0\right\}.
\end{equation}
Writing \( E_{\mathrm{obj}}(\bm{s})=\braket{\bm{s}|H_{\mathrm{obj}}|\bm{s}} \), we define the minimum feasible energy \( E_0 \)  and the set of optimal feasible configurations \( \mathcal{G} \)  as
\begin{align}
  E_0         & = \min_{\bm{s}\in\mathcal{F}} E_{\mathrm{obj}}(\bm{s}),                  \\
  \mathcal{G} & = \left\{\bm{s}\in\mathcal{F} \mid E_{\mathrm{obj}}(\bm{s})=E_0\right\}.
\end{align}
The computational-basis states \( \ket{g} \), with \( g\in\mathcal{G} \), span the ground-state manifold of \( H_{\mathrm{obj}} \) restricted to the feasible sector, with degeneracy \( d=|\mathcal{G}| \).
When discussing sampling, we refer to these computational-basis states as ground states.

For QA, the unnormalized problem and driver Hamiltonians are
\begin{align}
  H_{\mathrm{c}}^{\mathrm{QA}} & = H_{\mathrm{obj}}+\mu S_z^2, \\
  H_{\mathrm{q,X}}             & = -\sum_{i=1}^N\sigma_i^x,
\end{align}
where \( \mu > 0 \) is a penalty coefficient.
The penalty coefficient \( \mu \) must be large enough to ensure that the ground-state configurations of \( H_{\mathrm{c}}^{\mathrm{QA}} \) are exactly those in \( \mathcal{G} \).
However, an excessively large penalty coefficient can degrade annealing performance~\cite{lucas2014ising,yin2024penalty,ide2025extending}.
Here, we write \( \mu=\mu_{\mathrm{opt}}+\mu_{+} \), where \( \mu_{\mathrm{opt}} \) is given by
\begin{equation}\label{eq:penalty-threshold}
  \mu_{\mathrm{opt}} = \max_{\bm{s}\notin\mathcal{F}} \frac{E_0-E_{\mathrm{obj}}(\bm{s})}{\left(\sum_{i=1}^{N} s_i\right)^2}.
\end{equation}
The value of \( \mu_{\mathrm{opt}} \) is determined by exact enumeration of the diagonal objective and penalty energies.
At this threshold, the lowest feasible and infeasible energies coincide.
A strictly positive additional penalty \( \mu_{+} \) ensures that the ground-state configurations of \( H_{\mathrm{c}}^{\mathrm{QA}} \) are exactly those in \( \mathcal{G} \).
Thus, \( \mu_{\mathrm{opt}} \) denotes a feasibility threshold, rather than a penalty optimized for annealing performance or sampling uniformity.

For CQA, we use
\begin{align}
  H_{\mathrm{c}}^{\mathrm{CQA}} & = H_{\mathrm{obj}},                  \\
  H_{\mathrm{q,XY}}             & = -\frac{1}{2}\sum_{1\leq i<j\leq N}
  \left(\sigma_i^x\sigma_j^x+\sigma_i^y\sigma_j^y\right).
  \label{eq:xy-driver}
\end{align}
Because both \( H_{\mathrm{obj}} \) and \( H_{\mathrm{q,XY}} \) commute with \( S_z \), a state initialized in the \( S_z=0 \) sector remains in that sector throughout the evolution, and no penalty is needed.
The CQA target is therefore the ground-state manifold of \( H_{\mathrm{obj}} \) restricted to this sector.

\subsection{Annealing protocol and numerical setup}
For each method \( \alpha\in\{\mathrm{QA},\mathrm{CQA}\} \), we denote the quantum state at time \( t \) during annealing by \( \ket{\psi_{\alpha}(t)} \).
We obtain its time evolution by solving the closed-system Schr\"odinger equation, with \( \hbar=1 \), for the Hamiltonian
\begin{equation}
  H^{\alpha}(t)=A(t)H_{\mathrm{q}}^{\alpha}
  +\frac{B(t)}{\Lambda_{\alpha}}H_{\mathrm{c}}^{\alpha},
  \qquad 0\leq t\leq T,
  \label{eq:annealing-hamiltonian}
\end{equation}
where \( \Lambda_{\alpha} \) is a nonzero normalization constant for the problem Hamiltonian, defined below.
We use the linear schedules
\begin{equation}
  A(t)=1-\frac{t}{T},\qquad B(t)=\frac{t}{T},
\end{equation}
where \( T \) is the annealing time.
We normalize each problem Hamiltonian so that the largest absolute pairwise coupling is unity.
Specifically, writing
\begin{equation}
  H_{\mathrm{c}}^{\alpha}=C_{\alpha}I +\sum_{1\leq i<j\leq N}J_{ij}^{\alpha}\sigma_i^z\sigma_j^z,
\end{equation}
where \( C_{\alpha} \) is a constant energy offset, we set \( \Lambda_{\alpha}=\max_{1\leq i<j\leq N}|J_{ij}^{\alpha}| \) and use \( H_{\mathrm{c}}^{\alpha}/\Lambda_{\alpha} \) in the annealing Hamiltonian.
The driver Hamiltonians retain the coefficients specified above.
Annealing times reported here depend on these driver coefficients, the problem-energy normalization, and the schedules.
They do not provide a universal speed comparison between QA and CQA.
All penalty values and objective-energy differences reported below are given before this normalization.

For QA, we take as the initial state the unique ground state of the transverse-field driver \( H_{\mathrm{q,X}} \).
For CQA, we take as the initial state the unique ground state of the XY driver \( H_{\mathrm{q,XY}} \) within the feasible sector \( S_z=0 \).
These states are
\begin{align}
  \ket{\psi_{\mathrm{QA}}(0) }  & = \frac{1}{\sqrt{2^N}} \sum_{\bm{s}\in\{-1,+1\}^N}\ket{\bm{s}},     \\
  \ket{\psi_{\mathrm{CQA}}(0) } & = \binom{N}{N/2}^{-1/2} \sum_{\bm{s} \in \mathcal{F}} \ket{\bm{s}}.
\end{align}
The QA sum runs over all \( 2^N \) spin configurations, whereas the CQA sum runs over the \( \binom{N}{N/2} \) feasible configurations in \( \mathcal{F} \).

We use 100 random GBP instances for each of \( N=6,8,10,12 \), generated from the Erd\H{o}s--R\'enyi model with edge probability \( 0.8 \)~\cite{erdos1959random,gilbert1959random} and unit weights on present edges.
All graphs used in this study are connected.
Since each optimal partition is represented by two spin configurations related by global spin inversion, we retain only instances with at least four optimal feasible computational-basis states, corresponding to at least two distinct optimal partitions.
The same instances are compared across methods and penalties.
We sample annealing times on a logarithmic grid over \( 1\leq T\leq10^5 \) and additional penalties in the range \( 0.01\leq\mu_{+}\leq10 \).
The values displayed in each comparison are specified in the figures.

We integrate the Schr\"odinger equation in the full \( 2^N \)-dimensional Hilbert space using QuTiP~\cite{qutip5,johansson2013qutip,johansson2012qutip}.
The integration uses QuTiP's default Adams method, namely the variable-order (up to 12), variable-step Adams--Moulton integrator ZVODE~\cite{brown1989vode} accessed through SciPy~\cite{virtanen2020scipy}, with relative and absolute tolerances of \( 10^{-10} \) and \( 10^{-12} \), respectively.
These tolerances are considerably tighter than QuTiP's defaults (\( 10^{-6} \) and \( 10^{-8} \)) and are used to reduce numerical integration errors at long annealing times.

\subsection{Sampling observables}\label{subsec:sampling-observables}
For the final state \( \ket{\psi_{\alpha}(T)} \), the ground-state probability is
\begin{equation}
  P_{\mathrm{GS}}^{\alpha}
  =\sum_{g\in\mathcal{G}}|\braket{g|\psi_{\alpha}(T)}|^2.
\end{equation}
To characterize sampling among ground states, we define the conditional ground-state distribution, namely the probability of observing each ground state given that a ground state is obtained, as
\begin{equation}
  p_{\alpha}(g)=
  \frac{|\braket{g|\psi_{\alpha}(T)}|^2}{P_{\mathrm{GS}}^{\alpha}},
  \qquad g\in\mathcal{G},
  \label{eq:conditional-distribution}
\end{equation}
for \( P_{\mathrm{GS}}^{\alpha}>0 \).

For any two probability distributions \( p \) and \( q \) over \( \mathcal{G} \), we quantify their difference using the Jensen--Shannon (JS) divergence~\cite{lin1991divergence},
\begin{align}
  D_{\mathrm{JS}}(p,q) & =\frac{1}{2}D_{\mathrm{KL}}(p,m)+\frac{1}{2}D_{\mathrm{KL}}(q,m), \\
  m                    & =\frac{p+q}{2},                                                   \\
  D_{\mathrm{KL}}(p,q) & =\sum_{g\in\mathcal{G}}p(g)\log_2\frac{p(g)}{q(g)},
\end{align}
where \( m \) is the mixture distribution of \( p \) and \( q \), \(D_{\mathrm{KL}}\) is the Kullback--Leibler (KL) divergence~\cite{kullback1951information}, and terms with \( p(g)=0 \) contribute zero in the KL divergence.
Unlike the KL divergence, the JS divergence is symmetric in \( p \) and \( q \) and remains finite when one distribution assigns zero probability to a ground state to which the other assigns nonzero probability, as can occur under strongly biased sampling.
Its square root is a metric on probability distributions~\cite{endres2003new}, and \( D_{\mathrm{JS}} \) serves as a bounded measure of the distance between two distributions.
With this logarithm convention, \( 0\leq D_{\mathrm{JS}}\leq1 \), and the divergence vanishes if and only if the distributions coincide.
Let \( u(g)=1/d \) denote the uniform distribution over \( \mathcal{G} \).
We evaluate \( D_{\mathrm{JS}}(p_{\alpha},u) \) to measure sampling bias and \( D_{\mathrm{JS}}(p_{\mathrm{QA}},p_{\mathrm{CQA}}) \) to compare the methods at the same \( T \).
Averages are taken over the per-instance observables, including the per-instance divergences.

\section{Results}\label{sec:results}
We first characterize ground-state probability and sampling bias in QA and CQA, and examine when the two methods produce similar ground-state distributions.
We then use degenerate perturbation theory to explain their agreement at large penalties.
Finally, we examine the penalty dependence of distributional agreement and its consistency with the perturbative analysis.

\subsection{Ground-state probability and sampling bias}
\begin{figure*}[t!]
  \centering
  \includegraphics[width=\linewidth]{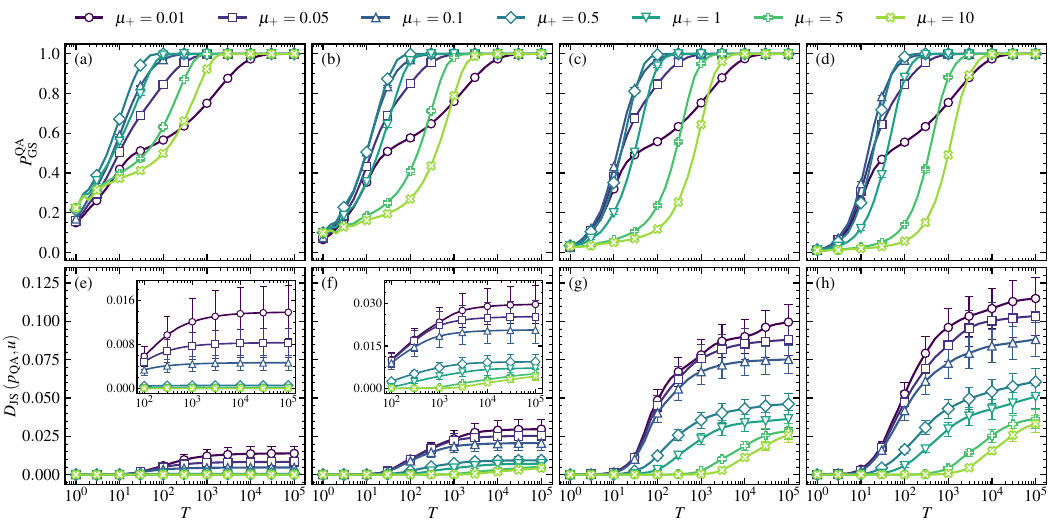}
  \caption{
    (a)--(d) Mean ground-state probability \( P_{\mathrm{GS}}^{\mathrm{QA}} \) and (e)--(h) mean divergence \( D_{\mathrm{JS}}(p_{\mathrm{QA}},u) \) from the uniform distribution, as functions of the annealing time \( T \).
    The distribution \( p_{\mathrm{QA}} \) is conditioned on obtaining a ground state.
    The insets in (e) and (f) show the same data for \( 10^2\leq T\leq10^5 \) on an enlarged vertical scale.
    Columns from left to right correspond to \( N = 6, 8, 10, 12 \), with both quantities averaged over 100 random GBP instances for each system size.
    Different colors and marker shapes distinguish values of the additional penalty coefficient \( \mu_{+} \).
    The error bars show the standard error of the mean.
    For clarity, markers and error bars are omitted at some computed annealing times, and the lines connect all computed data points.
  }
  \label{fig:qa-vs-uniform-js}
\end{figure*}

For both QA and CQA, we first evaluate the ground-state probability and the bias in the probability distribution over the degenerate ground states at each annealing time.
Figure~\ref{fig:qa-vs-uniform-js} shows the mean ground-state probability and the mean divergence from the uniform ground-state distribution for penalty-based QA.
The ground-state probability \( P_{\mathrm{GS}}^{\mathrm{QA}} \) approaches unity at shorter annealing times for intermediate values of the additional penalty \( \mu_{+} \) than for the smallest and largest penalties.
Increasing the penalty generally reduces the divergence from uniform sampling over the range examined.
This reduction persists at long annealing times, where the ground-state probability is close to unity.
The reduced divergence indicates that a larger penalty makes the conditional ground-state distribution more uniform.
For \( N=6 \), the divergence at the longest annealing time approaches zero as the penalty increases.
For \( N=8,10,12 \), residual bias remains, with larger mean divergences for larger systems.
This residual bias is consistent with the biased sampling reported for weighted GBP instances in Ref.~\cite{ide2026unfair}.
Because the residual bias grows with \( N \), sampling may become more strongly biased for systems larger than those simulated here.
In contrast, the divergence decreases with increasing penalty at all four sizes, which suggests that this trend also persists for larger \( N \).
Thus, the penalty affects both the ground-state probability and the relative probabilities within the ground-state manifold, and the two effects must be assessed separately.

\begin{figure}[t!]
  \centering
  \includegraphics[width=\linewidth]{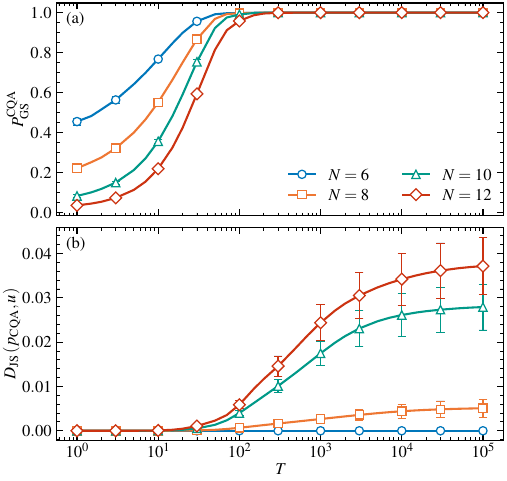}
  \caption{
    (a) Mean ground-state probability \( P_{\mathrm{GS}}^{\mathrm{CQA}} \) and (b) mean divergence \( D_{\mathrm{JS}}(p_{\mathrm{CQA}},u) \) from the uniform distribution, as functions of the annealing time \( T \).
    The distribution \( p_{\mathrm{CQA}} \) is conditioned on obtaining a ground state.
    Both quantities are averaged over 100 random GBP instances for each system size \( N=6,8,10,12 \), distinguished by colors and marker shapes.
    The error bars show the standard error of the mean.
    For clarity, markers and error bars are omitted at some computed annealing times, and the lines connect all computed data points.
  }
  \label{fig:cqa-vs-uniform-js}
\end{figure}

Figure~\ref{fig:cqa-vs-uniform-js} shows the corresponding CQA results for the same instances.
The mean ground-state probability approaches unity as the annealing time increases, with larger systems requiring longer times [Fig.~\ref{fig:cqa-vs-uniform-js}(a)].
The conditional ground-state distribution is nearly uniform at short times for all four sizes [Fig.~\ref{fig:cqa-vs-uniform-js}(b)].
For \( N=6 \), it remains nearly uniform throughout the simulated annealing times.
For \( N=8,10,12 \), however, the divergence increases with \( T \) and remains nonzero at the longest times.
In particular, the final distribution continues to change as \(T\) is increased, even when the ground-state probability is already close to unity.
As in QA, the divergence at the longest times grows with \( N \), so the sampling bias may become larger for systems larger than those simulated here.
CQA ensures feasibility, but the observed bias shows that it does not ensure uniform sampling among the optimal feasible states.
In Figs.~\ref{fig:qa-vs-uniform-js} and~\ref{fig:cqa-vs-uniform-js}, most divergences change little between \( T=10^4 \) and \( 10^5 \), whereas those for QA with larger \( \mu_{+} \) are still increasing at \( T=10^5 \).
As discussed in Secs.~\ref{subsec:perturbation-theory-analysis} and~\ref{subsec:comparison-and-penalty}, in the adiabatic limit of long annealing times, the divergence for QA is expected to approach a fixed value as \( \mu_{+} \) increases.

\subsection{Agreement between QA and CQA distributions}
\begin{figure*}[t!]
  \centering
  \includegraphics[width=\linewidth]{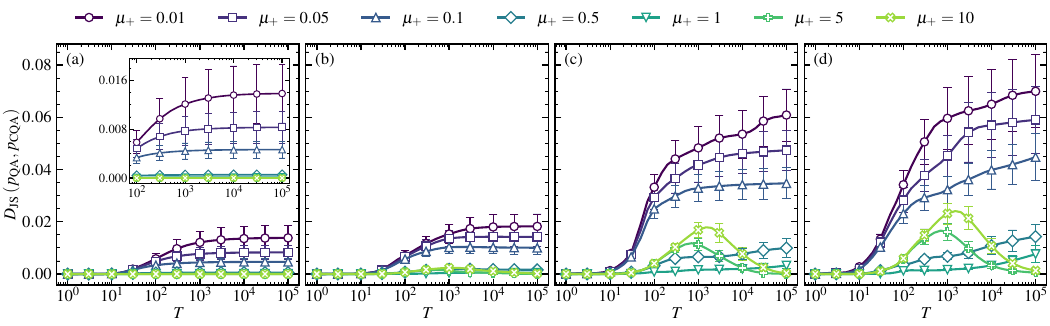}
  \caption{
    Mean divergence \( D_{\mathrm{JS}}(p_{\mathrm{QA}},p_{\mathrm{CQA}}) \) between the conditional ground-state distributions at the same annealing time \( T \), averaged over 100 instances per size.
    Colors and marker shapes distinguish the additional penalty \( \mu_{+} \).
    The error bars show the standard error of the mean.
    From left to right, the columns correspond to the system size \( N = 6, 8, 10, 12 \).
    For clarity, markers and error bars are omitted at some computed annealing times, and the lines connect all computed data points.
  }
  \label{fig:qa-vs-cqa-complete-js}
\end{figure*}

The residual bias in both methods raises the question of whether they can produce the same biased ground-state distribution.
Figure~\ref{fig:qa-vs-cqa-complete-js} compares their conditional distributions at the same annealing time.
At short times, the mean divergence is small for all values of \( \mu_{+} \) shown.
For sufficiently short annealing times, the states remain close to their initial states, which assign equal amplitudes to all computational-basis states in QA and all feasible states in CQA.
Both therefore yield nearly uniform distributions when conditioned on the ground-state manifold.
However, their total ground-state probabilities are far from unity and can differ between the two methods.

As the annealing time increases, the two conditional distributions separate because the sampling bias develops differently in the two methods.
For \(N=8, 10, 12\), the mean CQA divergence from uniform sampling continues to increase as \(T\) is increased, even when the mean ground-state probability is already close to unity.
In QA, the dependence of the mean divergence on \(T\) varies with the additional penalty \(\mu_{+}\) [Fig.~\ref{fig:qa-vs-uniform-js}(f)--(h)].
For the larger penalties, QA requires longer annealing times than CQA before its distribution becomes biased.
At intermediate annealing times, the CQA distribution is therefore already biased, whereas the QA distribution is still closer to uniform.
As a result, the divergence between the two methods for these penalties reaches a maximum at intermediate annealing times.
It then decreases and becomes small at the longest times.
For \( \mu_{+}=5 \) and \( 10 \), the QA divergences from uniform sampling are still increasing at \( T=10^5 \) but are already close to the CQA values, which suggests that they approach the CQA values at longer times.
For \( \mu_{+}\leq1 \), the mean divergence between the two methods shows no sign of the decrease seen for the larger penalties and is still increasing slowly at \( T=10^5 \) for \( N=10,12 \), which suggests that it remains nonzero at longer times.
For these small penalties, the divergence at \( T=10^5 \) also grows with \( N \), whereas for \( \mu_{+}=5 \) and \( 10 \) it becomes small at the longest times for all four sizes.
This suggests that, as \( N \) increases, the difference between the QA and CQA distributions widens for small penalties, while the two distributions still approach each other for large penalties.
These observations show that penalty-based QA approximately reproduces the CQA sampling bias on average for \( \mu_{+}=5,10 \) at \( T=10^5 \).
Agreement between the two methods does not imply fair sampling.

\subsection{Correspondence between effective Hamiltonians at large penalty}\label{subsec:perturbation-theory-analysis}
To explain the distributional agreement at large penalties in Fig.~\ref{fig:qa-vs-cqa-complete-js}, we use degenerate perturbation theory to obtain the effective Hamiltonians of the two methods within the degenerate ground-state manifold near the end of annealing and show that they correspond to each other.
For \( t>0 \), dividing Eq.~\eqref{eq:annealing-hamiltonian} by \( B(t)/\Lambda_{\alpha} \) gives the rescaled Hamiltonian for the instantaneous eigenvalue problem
\begin{equation}\label{eq:rescaled-hamiltonian}
  \widetilde{H}^{\alpha}=H_{\mathrm{c}}^{\alpha} +\Gamma_{\alpha}H_{\mathrm{q}}^{\alpha}, \qquad \Gamma_{\alpha}=\frac{A(t)\Lambda_{\alpha}}{B(t)}.
\end{equation}
We apply degenerate perturbation theory in the regime where the driver term \(\Gamma_\alpha H_{\mathrm q}^\alpha\) is weak compared with the unperturbed excitation gaps out of the ground-state manifold. For CQA, these gaps are evaluated within the feasible sector.

For QA with \( \mu>\mu_{\mathrm{opt}} \), we denote by \( P_g=\sum_{g\in\mathcal{G}}\ket{g}\bra{g} \) the projector onto the ground-state manifold and by \( Q_g=I-P_g \) its orthogonal complement.
Degenerate perturbation theory gives the effective Hamiltonian~\cite{bravyi2011schrieffer,jordan2008perturbative}
\begin{equation}
  \widetilde{H}_{\mathrm{eff}}^{\mathrm{QA}} =E_0P_g+\Gamma_{\mathrm{QA}}H_{\mathrm{eff}}^{\mathrm{QA},(1)} + \Gamma_{\mathrm{QA}}^2H_{\mathrm{eff}}^{\mathrm{QA},(2)}+O(\Gamma_{\mathrm{QA}}^3),
\end{equation}
where
\begin{align}
  H_{\mathrm{eff}}^{\mathrm{QA},(1)} & = P_gH_{\mathrm{q,X}}P_g,                   \\
  H_{\mathrm{eff}}^{\mathrm{QA},(2)} & = -P_gH_{\mathrm{q,X}}Q_g
  \left[Q_g(H_{\mathrm{c}}^{\mathrm{QA}}-E_0I)Q_g\right]^{-1}\notag                \\
                                     & \hspace{25mm}\times Q_gH_{\mathrm{q,X}}P_g.
  \label{eq:qa-second-order}
\end{align}
The inverse of \( Q_g(H_{\mathrm{c}}^{\mathrm{QA}}-E_0I)Q_g \) acts only on the range of \( Q_g \).
A single-spin flip changes \( S_z \) from zero to \( \pm2 \), so \( P_gH_{\mathrm{q,X}}P_g=0 \).
Thus, the first-order QA term vanishes within the ground-state manifold.
The leading nonzero interaction arises at second order through virtual single-spin flips to infeasible states.
This mechanism is analogous to that underlying the effective interactions appearing in perturbation theory for the half-filled Hubbard model~\cite{takahashi1977half,macdonald1988t}.
For an intermediate state \( m \) reached from \( g\in\mathcal{G} \) by one spin flip, the excitation energy is
\begin{equation}
  \begin{aligned}
    E_m-E_0                      & =4\mu+\Delta E_{\mathrm{obj}}(m,g), \\
    \Delta E_{\mathrm{obj}}(m,g) & =E_{\mathrm{obj}}(m)-E_0.
  \end{aligned}
  \label{eq:virtual-excitation}
\end{equation}

\begin{figure*}[t]
  \centering
  \includegraphics[width=\linewidth]{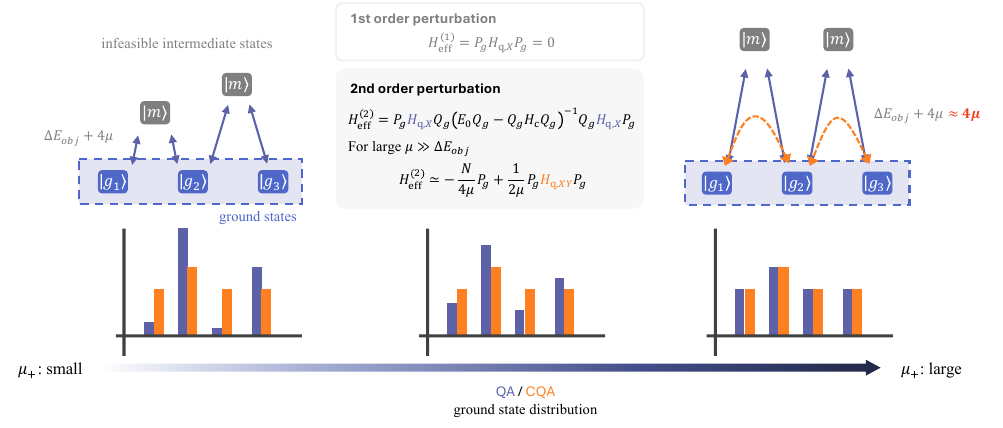}
  \caption{
    Schematic of the correspondence between the effective Hamiltonians of QA and CQA within the ground-state manifold.
    The horizontal arrow indicates increasing additional penalty \( \mu_{+} \).
    In QA, a single-spin flip takes a ground state \( \ket{g_k} \) to an infeasible intermediate state \( \ket{m} \) with excitation energy \( \Delta E_{\mathrm{obj}}+4\mu \) [Eq.~\eqref{eq:virtual-excitation}].
    The central boxes show the first- and second-order terms of the QA effective Hamiltonian.
    The first-order term vanishes because a single flip leaves the ground-state manifold, so the leading term arises at second order through virtual flips to these intermediate states.
    For small \( \mu_{+} \) (left), the excitation energies differ among the intermediate states.
    For large \( \mu_{+} \) (right), they can all be approximated by \( 4\mu \), and the second-order processes act as spin exchanges between ground states (dashed arrows) described by \( P_gH_{\mathrm{q,XY}}P_g \), the same operator that appears at first order in CQA.
    The bar charts show schematic ground-state distributions of QA (indigo) and CQA (orange), which approach each other as \( \mu_{+} \) increases.
  }
  \label{fig:perturbation-schematic}
\end{figure*}

If \( 4\mu\gg|\Delta E_{\mathrm{obj}}(m,g)| \) for all such intermediate states, their denominators may be replaced by \( 4\mu \).
Using \( P_gH_{\mathrm{q,X}}P_g=0 \), we then obtain
\begin{equation}
  H_{\mathrm{eff}}^{\mathrm{QA},(2)}\simeq
  -\frac{1}{4\mu}P_gH_{\mathrm{q,X}}^2P_g.
\end{equation}
Since
\begin{equation}
  H_{\mathrm{q,X}}^2=NI+2\sum_{1\leq i<j\leq N}\sigma_i^x\sigma_j^x,
\end{equation}
and the fixed-magnetization projection removes pair creation and annihilation, we have
\begin{equation}
  P_g\sigma_i^x\sigma_j^xP_g=
  \frac{1}{2}P_g\left(\sigma_i^x\sigma_j^x+\sigma_i^y\sigma_j^y\right)P_g
  \quad (i\ne j).
\end{equation}
Consequently, with the XY normalization in Eq.~\eqref{eq:xy-driver},
\begin{align}
  P_gH_{\mathrm{q,X}}^2P_g
   & =NP_g-2P_gH_{\mathrm{q,XY}}P_g, \\
  H_{\mathrm{eff}}^{\mathrm{QA},(2)}
   & \simeq-\frac{N}{4\mu}P_g
  +\frac{1}{2\mu}P_gH_{\mathrm{q,XY}}P_g.
  \label{eq:large-penalty-operator}
\end{align}
The first term in Eq.~\eqref{eq:large-penalty-operator} is a common energy shift, while the second determines the leading splitting within the ground-state manifold.
For example, consider two ground states \(\ket{g}\) and \(\ket{g'}\) that differ only by exchanging two opposite spins. The two virtual paths, each involving two single-spin flips, give a matrix element \(-2/(4\mu)=-1/(2\mu)\).
This equals \([1/(2\mu)] \braket{g'|H_{\mathrm{q,XY}}|g}\).
Substituting this second-order term into the QA effective Hamiltonian gives
\begin{equation}
  \widetilde{H}_{\mathrm{eff}}^{\mathrm{QA}}\simeq
  \left(E_0-\frac{N\Gamma_{\mathrm{QA}}^2}{4\mu}\right)P_g
  +\frac{\Gamma_{\mathrm{QA}}^2}{2\mu}P_gH_{\mathrm{q,XY}}P_g.
  \label{eq:effective-xy-correspondence}
\end{equation}
In CQA, the projected XY operator already acts within the feasible sector, so it appears at first order as
\begin{equation}
  \widetilde{H}_{\mathrm{eff}}^{\mathrm{CQA}}
  =E_0P_g+\Gamma_{\mathrm{CQA}}P_gH_{\mathrm{q,XY}}P_g+O(\Gamma_{\mathrm{CQA}}^2).
\end{equation}
Thus the effective interaction coefficients correspond as
\begin{equation}\label{eq:effective-coupling}
  \Gamma_{\mathrm{CQA}}\;\longleftrightarrow\;\frac{\Gamma_{\mathrm{QA}}^2}{2\mu}.
\end{equation}
Figure~\ref{fig:perturbation-schematic} illustrates this correspondence schematically.

For the balance constraint considered here, both annealing Hamiltonians commute with the global spin-inversion operator \( R=\prod_{i=1}^{N}\sigma_i^x \), and both initial states are even under \( R \)~\cite{tanaka2010nonmonotonic}.
Both evolutions therefore remain in the \( R=+1 \) sector, with CQA additionally restricted to \( S_z=0 \).
State selection must therefore be examined within the \( R=+1 \) subspace of the ground-state manifold, subject to any further restrictions imposed by symmetries shared by the Hamiltonians and initial states.
When the projected XY operator \( P_gH_{\mathrm{q,XY}}P_g \) has a unique lowest eigenstate within this symmetry-restricted subspace, the correspondence in Eq.~\eqref{eq:effective-coupling} identifies the same leading state selection for the two methods in the large-penalty limit.
That state need not assign equal amplitudes to all configurations in \( \mathcal{G} \), so the correspondence also allows a common sampling bias.
If the lowest eigenvalue remains degenerate within this subspace, the common leading operator alone does not determine the selected state.
The terms that may lift this degeneracy include finite-penalty corrections retained in the second-order QA term in Eq.~\eqref{eq:qa-second-order} and higher-order terms in the respective driver strengths.
These statements concern instantaneous eigenstates near the end of annealing.
Final distributional agreement additionally depends on whether the finite-time evolution follows the relevant state.

This correspondence between the effective Hamiltonians extends to an equal-weight constraint \( \sum_i s_i=C \) with penalty \( \mu(S_z-CI)^2 \), where the integer \( C \) is the target magnetization, provided the feasible sector is nonempty, all ground states of the penalized problem Hamiltonian are feasible, and the same perturbative conditions hold.
A single flip still incurs penalty energy \( 4\mu \), and the fixed-magnetization projection again retains spin exchanges.

\subsection{Penalty dependence of distributional agreement}\label{subsec:comparison-and-penalty}
\begin{figure}
  \centering
  \includegraphics[width=\linewidth]{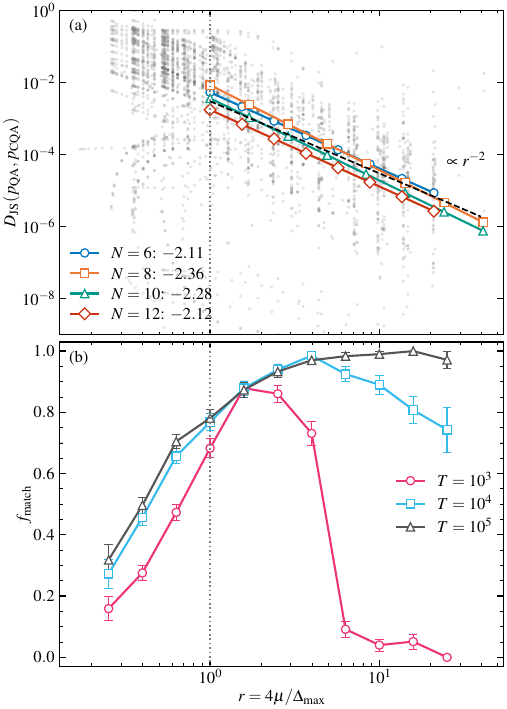}
  \caption{
    Agreement between the conditional ground-state distributions of QA and CQA as a function of $r=4\mu/\Delta_{\max}$.
    (a) JS divergence $D_{\mathrm{JS}}(p_{\mathrm{QA}},p_{\mathrm{CQA}})$ at $T=10^5$ for $N=6,8,10,12$.
    Each faint dot represents one instance--penalty pair.
    For each $N$, the slope of the solid line is the median of the per-instance log--log slopes fitted over $r\geq1$ using points with $D_{\mathrm{JS}}>10^{-10}$.
    This median slope is given in the legend.
    The symbols on each line are shown only for visibility.
    The median includes instances for which a slope can be estimated from the retained points.
    The dashed line guides the eye with an $r^{-2}$ dependence.
    (b) Fraction $f_{\mathrm{match}}$ of instance--penalty pairs satisfying $P_{\mathrm{GS}}^{\mathrm{QA}}>0.99$, $P_{\mathrm{GS}}^{\mathrm{CQA}}>0.99$, and $D_{\mathrm{JS}}<0.01$ for $N=12$ at $T=10^3,10^4,10^5$.
    Each plotted point gives this fraction within a logarithmic bin containing at least 10 instance--penalty pairs, and error bars show nominal binomial standard errors.
    Points with $D_{\mathrm{JS}}\leq10^{-10}$ are omitted from the plot and fits in (a), but are included in (b).
    Dotted vertical lines mark $r=1$.
  }
  \label{fig:qa-cqa-match-pt}
\end{figure}

The large-penalty approximation suggests an instance-dependent scale for distributional agreement.
We define
\begin{equation}
  \Delta_{\max}=\max_{g,m}|\Delta E_{\mathrm{obj}}(m,g)|,
  \qquad r=\frac{4\mu}{\Delta_{\max}},
  \label{eq:penalty-ratio}
\end{equation}
where the maximum is taken over all \(g\in\mathcal{G}\) and all spin configurations \(m\) obtained from \(g\) by flipping exactly one spin.
In all instances studied, \( \Delta_{\max} \) is nonzero, so \( r \) is well defined.
The condition \( r\gg1 \) controls the replacement of the virtual excitation energies by \( 4\mu \).

Figure~\ref{fig:qa-cqa-match-pt}(a) shows the JS divergence \( D_{\mathrm{JS}}(p_{\mathrm{QA}},p_{\mathrm{CQA}}) \) between the simulated conditional ground-state distributions of QA and CQA at \( T=10^5 \) as a function of \( r \).
Despite substantial scatter among instance--penalty pairs, the divergence decreases approximately as \( r^{-2} \) for \( r>1 \), with similar trends across the four system sizes.
To quantify this trend, we fit a straight line to \( \log_{10}D_{\mathrm{JS}} \) versus \( \log_{10}r \) for each instance using points with \( r\geq1 \) and \( D_{\mathrm{JS}}>10^{-10} \), and take the median of the resulting slopes for each \( N \).
Instances for which a slope cannot be estimated from the retained points do not enter this median.
The median slopes range from \( -2.36 \) to \( -2.11 \), as listed in Fig.~\ref{fig:qa-cqa-match-pt}(a).
Expanding the reciprocals of the exact denominators \( 4\mu+\Delta E_{\mathrm{obj}}(m,g) \) of Eq.~\eqref{eq:qa-second-order} in powers of \( \Delta E_{\mathrm{obj}}(m,g)/(4\mu) \) gives relative corrections to the large-penalty operator of order \( r^{-1} \).
For a fixed instance with a nondegenerate lowest projected XY eigenstate in the relevant symmetry sector, sufficiently small corrections produce eigenstate changes of the same order.
When the resulting probabilities vary smoothly on a common positive support, the JS divergence is quadratic in their difference to leading order, giving an \( r^{-2} \) dependence.
The slightly steeper fitted slopes may reflect finite-\( r \) corrections, although finite-time effects and the fitting procedure can also contribute.
The perturbative argument accounts for an approximately inverse-square trend under the stated conditions, but does not establish a universal scaling law for every instance or annealing time.

To assess distributional agreement when both methods have high ground-state probability, we count an instance as a match at a given \( T \) and \( \mu_{+} \) when
\begin{equation}
  \begin{gathered}
    P_{\mathrm{GS}}^{\mathrm{QA}}>0.99,\qquad
    P_{\mathrm{GS}}^{\mathrm{CQA}}>0.99,\\
    D_{\mathrm{JS}}(p_{\mathrm{QA}},p_{\mathrm{CQA}})<0.01.
  \end{gathered}
  \label{eq:matching-criterion}
\end{equation}
The probability requirements distinguish this criterion from agreement between conditional distributions alone, which can occur even when ground-state probabilities are low.
Figure~\ref{fig:qa-cqa-match-pt}(b) shows the resulting agreement fraction \( f_{\mathrm{match}} \) as a function of \( r \).
For \( N=12 \) and \( T=10^5 \), the fraction of sampled instance--penalty pairs satisfying the criterion rises to about \( 0.95 \) or above for \( r\gtrsim5 \), with some bins reaching unity.
Thus, \( r \) of order \( 5\text{--}10 \) is an empirical scale for agreement in this ensemble at the specified time and tolerance.
Equivalently, \( \mu\simeq(5\text{--}10)\Delta_{\max}/4 \).
This range is not an instance-independent bound for exact agreement, and the plotted fractions pool instance--penalty pairs rather than unique instances.

At shorter times, increasing \( r \) does not consistently improve the matching fraction.
At \( T=10^3 \), the fraction peaks near \( r\simeq2 \) and then falls sharply.
At \( T=10^4 \), it remains high over a broader range before decreasing at large \( r \).
These finite-time observations show that improving the large-penalty operator approximation does not by itself ensure a higher matching fraction at a given annealing time.
To identify what limits agreement at each \( r \), we examined which conditions in Eq.~\eqref{eq:matching-criterion} are violated in the unmatched pairs.
For \( N=12 \) and \( T=10^3\text{--}10^5 \), almost all unmatched pairs violate the QA ground-state probability condition or the divergence condition, rather than the CQA ground-state probability condition.
This is consistent with the mean CQA ground-state probability in Fig.~\ref{fig:cqa-vs-uniform-js}(a).
At small \( r \), most unmatched pairs have \( P_{\mathrm{GS}}^{\mathrm{QA}}>0.99 \) but \( D_{\mathrm{JS}}\geq0.01 \).
At large \( r \) and \( T\leq10^4 \), most unmatched pairs instead have \( P_{\mathrm{GS}}^{\mathrm{QA}}\leq0.99 \), consistent with the longer annealing times required for \( P_{\mathrm{GS}}^{\mathrm{QA}} \) to approach unity at the largest penalties in Fig.~\ref{fig:qa-vs-uniform-js}(d).
At \( T=10^5 \), almost all pairs satisfy \( P_{\mathrm{GS}}^{\mathrm{QA}}>0.99 \), and the remaining unmatched pairs at small \( r \) violate the divergence condition.

\section{Conclusion}\label{sec:conclusion}
QA and CQA can produce similar biased distributions over degenerate ground states, despite enforcing the constraint through different mechanisms.
We numerically simulated the closed-system dynamics of QA and CQA for 100 random GBP instances at each of \( N=6,8,10,12 \).
From these simulations, we evaluated the probability of reaching a ground state, the uniformity of the conditional ground-state distribution, and the distributional agreement between the two methods.
Increasing the penalty coefficient generally reduced the QA sampling bias, but residual bias remained for the larger systems at the longest simulated times.
CQA also exhibited nonuniform sampling, showing that feasibility and a high ground-state probability do not ensure fairness.
At sufficiently large penalties and long annealing times, the QA conditional ground-state distribution approached that of CQA in the studied ensembles, so the two methods produced similar sampling biases.

Degenerate perturbation theory explains why the two methods can approach the same distribution.
In the effective Hamiltonian of QA near the end of annealing, virtual single-spin flips to infeasible states generate a second-order interaction that, in the large-penalty approximation, is proportional to the CQA driver projected onto the feasible ground-state manifold, whereas the effective Hamiltonian of CQA contains the same projected operator at first order.
In the simulations at \( T = 10^5 \), the JS divergence between the conditional ground-state distributions of QA and CQA decreased approximately as \( r^{-2} \) for \( r > 1 \), where \( r=4\mu/\Delta_{\max} \).
This dependence is consistent with the perturbative argument in Sec.~\ref{subsec:comparison-and-penalty}.
Under the assumptions stated there, relative finite-penalty corrections of order \( r^{-1} \) to the effective interaction yield a divergence of order \( r^{-2} \).
The correspondence between the effective Hamiltonians specifies the same leading state selection for the two methods when the projected XY operator \( P_gH_{\mathrm{q,XY}}P_g \) has a unique lowest eigenstate within the dynamically accessible symmetry sector.
This correspondence does not imply identical finite-time dynamics.
At finite annealing times, increasing the penalty did not always improve the matching fraction, mainly because QA required longer annealing times to reach a ground-state probability close to unity at large penalties.

Beyond a comparison of two annealing protocols, the present results provide a concrete example of distinct microscopic mechanisms generating the same leading interaction within a degenerate manifold.
This raises the broader question of which features of selection by quantum fluctuations depend on the effective transition structure and which depend on the constraint implementation.
Extending this comparison to other constrained and frustrated systems could clarify when higher-order virtual processes or thermal fluctuations preserve common state selection and when they cause the implementations to favor different states.
For QA, this perspective suggests treating constraint implementation as a means of engineering quantum fluctuations rather than merely excluding infeasible configurations.
A natural next step is to determine whether controllable changes to the driver and virtual excitation energies can reshape the distribution of optimal feasible configurations.
The goal would be to transition from explaining a shared sampling bias to identifying which distributions can be realized with the available interactions.
The same questions motivate comparisons between penalty encodings and constraint-preserving mixers in gate-based quantum optimization.
Digitized annealing is a good starting point when time-discretization errors are controlled.
However, extending the correspondence to the quantum approximate optimization algorithm at finite depth~\cite{farhi2014quantum,hadfield2019quantum} requires accounting for pulse ordering and variational parameters.
Identifying where the correspondence survives and breaks down could inform the design of mixers and Hamiltonians for sampling objectives beyond energy minimization.

\acknowledgments
This work was partially supported by the Japan Society for the Promotion of Science (JSPS) KAKENHI (Grant Number JP23H05447), the Council for Science, Technology, and Innovation (CSTI) through the Cross-ministerial Strategic Innovation Promotion Program (SIP), ``Promoting the application of advanced quantum technology platforms to social issues'' (Funding agency: QST), and the Japan Science and Technology Agency (JST) (Grant Number JPMJPF2221).
The computations in this work were partially performed using the facilities of the Supercomputer Center, the Institute for Solid State Physics, The University of Tokyo.
Part of the work by S.I. was conducted through activities in the 2026 NICT Quantum Camp (NQC), organized by the National Institute of Information and Communications Technology (NICT), Japan.
S.I. gratefully acknowledges the financial support provided through the program.
S.T. wishes to express their gratitude to the World Premier International Research Center Initiative (WPI), MEXT, Japan, for their support of the Human Biology-Microbiome-Quantum Research Center (Bio2Q).

\bibliography{main}

@PREAMBLE{
 "\providecommand{\noopsort}[1]{}" 
 # "\providecommand{\singleletter}[1]{#1}%" 
}

@article{kadowaki1998quantum,
  title     = {Quantum annealing in the transverse {Ising} model},
  author    = {Kadowaki, Tadashi and Nishimori, Hidetoshi},
  journal   = {Phys. Rev. E},
  volume    = {58},
  number    = {5},
  pages     = {5355},
  year      = {1998},
  doi       = {10.1103/PhysRevE.58.5355},
  publisher = {APS}
}

@article{farhi2001quantum,
  title     = {A quantum adiabatic evolution algorithm applied to random instances of an {NP}-complete problem},
  author    = {Farhi, Edward and Goldstone, Jeffrey and Gutmann, Sam and Lapan, Joshua and Lundgren, Andrew and Preda, Daniel},
  journal   = {Science},
  volume    = {292},
  number    = {5516},
  pages     = {472--475},
  year      = {2001},
  doi       = {10.1126/science.1057726},
  publisher = {American Association for the Advancement of Science}
}

@article{farhi2014quantum,
  title   = {A quantum approximate optimization algorithm},
  author  = {Farhi, Edward and Goldstone, Jeffrey and Gutmann, Sam},
  journal = {arXiv:1411.4028},
  year    = {2014}
}

@article{hen2016quantum,
  title     = {Quantum annealing for constrained optimization},
  author    = {Hen, Itay and Spedalieri, Federico M},
  journal   = {Phys. Rev. Appl.},
  volume    = {5},
  number    = {3},
  pages     = {034007},
  year      = {2016},
  doi       = {10.1103/PhysRevApplied.5.034007},
  publisher = {APS}
}

@article{hadfield2019quantum,
  title     = {From the quantum approximate optimization algorithm to a quantum alternating operator ansatz},
  author    = {Hadfield, Stuart and Wang, Zhihui and O'Gorman, Bryan and Rieffel, Eleanor G and Venturelli, Davide and Biswas, Rupak},
  journal   = {Algorithms},
  volume    = {12},
  number    = {2},
  pages     = {34},
  year      = {2019},
  doi       = {10.3390/a12020034},
  publisher = {MDPI}
}

@inproceedings{ide2026unfair,
  title         = {Unfair Sampling of Quantum Annealing in Weighted Graph Bipartitioning Problems},
  author        = {Ide, Shunta and Tanaka, Shu},
  year          = {2026},
  booktitle     = {Proceedings of the 2026 IEEE International Conference on Quantum Computing and Engineering (QCE)},
  note          = {(to be published)},
  eprint        = {2604.11449},
  archiveprefix = {arXiv}
}

@article{matsuda2009ground,
  title   = {Ground-state statistics from annealing algorithms: quantum versus classical approaches},
  author  = {Matsuda, Yoshiki and Nishimori, Hidetoshi and Katzgraber, Helmut G.},
  journal = {New J. Phys.},
  volume  = {11},
  number  = {7},
  pages   = {073021},
  year    = {2009},
  doi     = {10.1088/1367-2630/11/7/073021}
}

@article{mandra2017exponentially,
  title   = {Exponentially Biased Ground-State Sampling of Quantum Annealing Machines with Transverse-Field Driving {Hamiltonians}},
  author  = {Mandr{\`a}, Salvatore and Zhu, Zheng and Katzgraber, Helmut G.},
  journal = {Phys. Rev. Lett.},
  volume  = {118},
  number  = {7},
  pages   = {070502},
  year    = {2017},
  doi     = {10.1103/PhysRevLett.118.070502}
}

@inproceedings{garey1974some,
  title     = {Some simplified {NP-complete} problems},
  author    = {Garey, Michael R and Johnson, David S and Stockmeyer, Larry},
  booktitle = {Proceedings of the Sixth Annual ACM Symposium on Theory of Computing},
  pages     = {47--63},
  year      = {1974},
  doi       = {10.1145/800119.803884}
}

@article{sotirov2018graph,
  title     = {Graph bisection revisited},
  author    = {Sotirov, Renata},
  journal   = {Ann. Oper. Res.},
  volume    = {265},
  number    = {1},
  pages     = {143--154},
  year      = {2018},
  doi       = {10.1007/s10479-017-2575-3},
  publisher = {Springer}
}

@article{lucas2014ising,
  title     = {Ising formulations of many {NP} problems},
  author    = {Lucas, Andrew},
  journal   = {Front. Phys.},
  volume    = {2},
  pages     = {5},
  year      = {2014},
  doi       = {10.3389/fphy.2014.00005},
  publisher = {Frontiers Media SA}
}

@article{yin2024penalty,
  title     = {Penalty Coefficient Adjustment Technique for Extended {Ising} Machines},
  author    = {Yin, Fang and Tamura, Hirotaka and Furue, Yuki and Watanabe, Yasuhiro},
  journal   = {IEEE Access},
  volume    = {12},
  pages     = {168303--168313},
  year      = {2024},
  doi       = {10.1109/ACCESS.2024.3494736},
  publisher = {IEEE}
}

@article{qutip5,
  title   = {{QuTiP 5}: The Quantum Toolbox in {Python}},
  author  = {
             Lambert, Neill and Gigu{\`e}re, Eric and Menczel, Paul and Li, Boxi and
             Hopf, Patrick and Su{\'a}rez, Gerardo and Gali, Marc and Lishman, Jake and
             Gadhvi, Rushiraj and Agarwal, Rochisha and Galicia, Asier and Shammah, Nathan and
             Nation, Paul and Johansson, J. R. and Ahmed, Shahnawaz and Cross, Simon and
             Pitchford, Alexander and Nori, Franco
             },
  journal = {Phys. Rep.},
  volume  = {1153},
  pages   = {1--62},
  year    = {2026},
  issn    = {0370-1573},
  doi     = {10.1016/j.physrep.2025.10.001},
  url     = {https://www.sciencedirect.com/science/article/pii/S0370157325002704}
}

@article{brown1989vode,
  title     = {{VODE}: A variable-coefficient {ODE} solver},
  author    = {Brown, Peter N and Byrne, George D and Hindmarsh, Alan C},
  journal   = {SIAM J. Sci. Stat. Comput.},
  volume    = {10},
  number    = {5},
  pages     = {1038--1051},
  year      = {1989},
  doi       = {10.1137/0910062},
  publisher = {SIAM}
}

@article{virtanen2020scipy,
  author  = {Virtanen, Pauli and Gommers, Ralf and Oliphant, Travis E. and
             Haberland, Matt and Reddy, Tyler and Cournapeau, David and
             Burovski, Evgeni and Peterson, Pearu and Weckesser, Warren and
             Bright, Jonathan and {van der Walt}, St{\'e}fan J. and
             Brett, Matthew and Wilson, Joshua and Millman, K. Jarrod and
             Mayorov, Nikolay and Nelson, Andrew R. J. and Jones, Eric and
             Kern, Robert and Larson, Eric and Carey, C J and
             Polat, {\.I}lhan and Feng, Yu and Moore, Eric W. and
             {VanderPlas}, Jake and Laxalde, Denis and Perktold, Josef and
             Cimrman, Robert and Henriksen, Ian and Quintero, E. A. and
             Harris, Charles R. and Archibald, Anne M. and
             Ribeiro, Ant{\^o}nio H. and Pedregosa, Fabian and
             {van Mulbregt}, Paul and {SciPy 1.0 Contributors}},
  title   = {{{SciPy} 1.0: Fundamental Algorithms for Scientific
             Computing in Python}},
  journal = {Nat. Methods},
  year    = {2020},
  volume  = {17},
  pages   = {261--272},
  adsurl  = {https://rdcu.be/b08Wh},
  doi     = {10.1038/s41592-019-0686-2}
}

@article{johansson2013qutip,
  title    = {{QuTiP 2}: A {Python} framework for the dynamics of open quantum systems},
  journal  = {Comput. Phys. Commun.},
  volume   = {184},
  number   = {4},
  pages    = {1234--1240},
  year     = {2013},
  issn     = {0010-4655},
  doi      = {10.1016/j.cpc.2012.11.019},
  url      = {https://www.sciencedirect.com/science/article/pii/S0010465512003955},
  author   = {J. R. Johansson and P. D. Nation and Franco Nori}
}

@article{johansson2012qutip,
  title    = {{QuTiP}: An open-source {Python} framework for the dynamics of open quantum systems},
  journal  = {Comput. Phys. Commun.},
  volume   = {183},
  number   = {8},
  pages    = {1760--1772},
  year     = {2012},
  issn     = {0010-4655},
  doi      = {10.1016/j.cpc.2012.02.021},
  url      = {https://www.sciencedirect.com/science/article/pii/S0010465512000835},
  author   = {J. R. Johansson and P. D. Nation and Franco Nori}
}

@article{lin1991divergence,
  title     = {Divergence measures based on the {Shannon} entropy},
  author    = {Lin, Jianhua},
  journal   = {IEEE Trans. Inf. Theory},
  volume    = {37},
  number    = {1},
  pages     = {145--151},
  year      = {1991},
  doi       = {10.1109/18.61115},
  publisher = {IEEE}
}

@article{endres2003new,
  title     = {A new metric for probability distributions},
  author    = {Endres, Dominik Maria and Schindelin, Johannes E},
  journal   = {IEEE Trans. Inf. Theory},
  volume    = {49},
  number    = {7},
  pages     = {1858--1860},
  year      = {2003},
  doi       = {10.1109/TIT.2003.813506},
  publisher = {IEEE}
}

@article{kullback1951information,
  title     = {On information and sufficiency},
  author    = {Kullback, Solomon and Leibler, Richard A},
  journal   = {Ann. Math. Stat.},
  volume    = {22},
  number    = {1},
  pages     = {79--86},
  year      = {1951},
  doi       = {10.1214/aoms/1177729694},
  publisher = {JSTOR}
}

@article{ide2025extending,
  title   = {Extending sample persistence variable reduction for constrained combinatorial optimization problems},
  author  = {Ide, Shunta and Kikuchi, Shuta and Tanaka, Shu},
  journal = {arXiv:2509.19280},
  year    = {2025}
}

@article{finnila1994quantum,
  title     = {Quantum annealing: A new method for minimizing multidimensional functions},
  author    = {Finnila, Aleta Berk and Gomez, Maria A and Sebenik, C and Stenson, Catherine and Doll, Jimmie D},
  journal   = {Chem. Phys. Lett.},
  volume    = {219},
  number    = {5--6},
  pages     = {343--348},
  year      = {1994},
  doi       = {10.1016/0009-2614(94)00117-0},
  publisher = {Elsevier}
}

@article{santoro2002theory,
  title     = {Theory of quantum annealing of an {Ising} spin glass},
  author    = {Santoro, Giuseppe E and Marton{\'a}k, Roman and Tosatti, Erio and Car, Roberto},
  journal   = {Science},
  volume    = {295},
  number    = {5564},
  pages     = {2427--2430},
  year      = {2002},
  doi       = {10.1126/science.1068774},
  publisher = {American Association for the Advancement of Science}
}

@article{kitai2020designing,
  title     = {Designing metamaterials with quantum annealing and factorization machines},
  author    = {Kitai, Koki and Guo, Jiang and Ju, Shenghong and Tanaka, Shu and Tsuda, Koji and Shiomi, Junichiro and Tamura, Ryo},
  journal   = {Phys. Rev. Res.},
  volume    = {2},
  number    = {1},
  pages     = {013319},
  year      = {2020},
  doi       = {10.1103/PhysRevResearch.2.013319},
  publisher = {APS}
}

@article{endo2022phase,
  title     = {A phase-field model by an {Ising} machine and its application to the phase-separation structure of a diblock polymer},
  author    = {Endo, Katsuhiro and Matsuda, Yoshiki and Tanaka, Shu and Muramatsu, Mayu},
  journal   = {Sci. Rep.},
  volume    = {12},
  number    = {1},
  pages     = {10794},
  year      = {2022},
  doi       = {10.1038/s41598-022-14735-4},
  publisher = {Nature Publishing Group UK London}
}

@article{inoue2022towards,
  title     = {Towards optimization of photonic-crystal surface-emitting lasers via quantum annealing},
  author    = {Inoue, Takuya and Seki, Yuya and Tanaka, Shu and Togawa, Nozomu and Ishizaki, Kenji and Noda, Susumu},
  journal   = {Opt. Express},
  volume    = {30},
  number    = {24},
  pages     = {43503--43512},
  year      = {2022},
  doi       = {10.1364/OE.476839},
  publisher = {Optica Publishing Group}
}

@article{xu2025quantum,
  title     = {Quantum annealing-assisted lattice optimization},
  author    = {Xu, Zhihao and Shang, Wenjie and Kim, Seongmin and Lee, Eungkyu and Luo, Tengfei},
  journal   = {npj Comput. Mater.},
  volume    = {11},
  number    = {1},
  pages     = {4},
  year      = {2025},
  doi       = {10.1038/s41524-024-01505-1},
  publisher = {Nature Publishing Group UK London}
}

@article{tanahashi2019application,
  title     = {Application of {Ising} machines and a software development for {Ising} machines},
  author    = {Tanahashi, Kotaro and Takayanagi, Shinichi and Motohashi, Tomomitsu and Tanaka, Shu},
  journal   = {J. Phys. Soc. Jpn.},
  volume    = {88},
  number    = {6},
  pages     = {061010},
  year      = {2019},
  doi       = {10.7566/JPSJ.88.061010},
  publisher = {The Physical Society of Japan}
}

@article{takahashi2026effectiveness,
  title     = {Effectiveness of cardinality-return weighted maximum independent set approach for financial portfolio optimization},
  author    = {Takahashi, Keita and Abe, Tetsuro and Nakamura, Yasuhito and Hidaka, Ryo and Kikuchi, Shuta and Tanaka, Shu},
  journal   = {J. Phys. Soc. Jpn.},
  volume    = {95},
  number    = {10},
  pages     = {104802},
  year      = {2026},
  doi       = {10.7566/JPSJ.95.104802},
  publisher = {The Physical Society of Japan}
}

@article{orus2019forecasting,
  title     = {Forecasting financial crashes with quantum computing},
  author    = {Or{\'u}s, Rom{\'a}n and Mugel, Samuel and Lizaso, Enrique},
  journal   = {Phys. Rev. A},
  volume    = {99},
  number    = {6},
  pages     = {060301},
  year      = {2019},
  doi       = {10.1103/PhysRevA.99.060301},
  publisher = {APS}
}

@article{bao2023ising,
  title     = {An {Ising}-machine-based solver of vehicle routing problem with balanced pick-up},
  author    = {Bao, Siya and Tawada, Masashi and Tanaka, Shu and Togawa, Nozomu},
  journal   = {IEEE Trans. Consum. Electron.},
  volume    = {70},
  number    = {1},
  pages     = {445--459},
  year      = {2024},
  doi       = {10.1109/TCE.2023.3335392},
  publisher = {IEEE}
}

@article{inoue2021traffic,
  title     = {Traffic signal optimization on a square lattice with quantum annealing},
  author    = {Inoue, Daisuke and Okada, Akihisa and Matsumori, Tadayoshi and Aihara, Kazuyuki and Yoshida, Hiroaki},
  journal   = {Sci. Rep.},
  volume    = {11},
  number    = {1},
  pages     = {3303},
  year      = {2021},
  doi       = {10.1038/s41598-021-82740-0},
  publisher = {Nature Publishing Group UK London}
}

@inproceedings{sahai2020estimating,
  title     = {Estimating the density of states of {Boolean} satisfiability problems on classical and quantum computing platforms},
  author    = {Sahai, Tuhin and Mishra, Anurag and Pasini, Jose Miguel and Jha, Susmit},
  booktitle = {Proceedings of the AAAI Conference on Artificial Intelligence},
  volume    = {34},
  pages     = {1627--1635},
  year      = {2020},
  doi       = {10.1609/aaai.v34i02.5524}
}

@article{konz2019uncertain,
  title     = {Uncertain fate of fair sampling in quantum annealing},
  author    = {K{\"o}nz, Mario S and Mazzola, Guglielmo and Ochoa, Andrew J and Katzgraber, Helmut G and Troyer, Matthias},
  journal   = {Phys. Rev. A},
  volume    = {100},
  number    = {3},
  pages     = {030303},
  year      = {2019},
  doi       = {10.1103/PhysRevA.100.030303},
  publisher = {APS}
}

@article{maruyama2026uniformity,
  title     = {Uniformity Bias in Ground-State Sampling Induced by Replica Alignment in Quantum {Monte Carlo} for Quantum Annealing},
  author    = {Maruyama, Naoki and Ohzeki, Masayuki and Tanaka, Kazuyuki},
  journal   = {J. Phys. Soc. Jpn.},
  volume    = {95},
  number    = {3},
  pages     = {034001},
  year      = {2026},
  doi       = {10.7566/JPSJ.95.034001},
  publisher = {The Physical Society of Japan}
}

@article{azinovic2017assessment,
  title   = {Assessment of quantum annealing for the construction of satisfiability filters},
  author  = {Azinovi{\'c}, Marlon and Herr, Daniel and Heim, Bettina and Brown, Ethan and Troyer, Matthias},
  journal = {SciPost Phys.},
  volume  = {2},
  number  = {2},
  pages   = {013},
  year    = {2017},
  doi     = {10.21468/SciPostPhys.2.2.013}
}

@article{zhang2017advantages,
  title     = {Advantages of unfair quantum ground-state sampling},
  author    = {Zhang, Brian Hu and Wagenbreth, Gene and Martin-Mayor, Victor and Hen, Itay},
  journal   = {Sci. Rep.},
  volume    = {7},
  number    = {1},
  pages     = {1044},
  year      = {2017},
  doi       = {10.1038/s41598-017-01096-6},
  publisher = {Nature Publishing Group UK London}
}

@article{valiant1979complexity,
  author  = {Valiant, Leslie G.},
  title   = {The Complexity of Enumeration and Reliability Problems},
  journal = {SIAM J. Comput.},
  volume  = {8},
  number  = {3},
  pages   = {410--421},
  year    = {1979},
  doi     = {10.1137/0208032}
}

@article{takahashi1977half,
  title   = {Half-filled {Hubbard} model at low temperature},
  author  = {Takahashi, Minoru},
  journal = {J. Phys. C: Solid State Phys.},
  volume  = {10},
  number  = {8},
  pages   = {1289},
  year    = {1977},
  doi     = {10.1088/0022-3719/10/8/031}
}

@article{macdonald1988t,
  title     = {{\( t/U \)} expansion for the {Hubbard} model},
  author    = {MacDonald, Allan H and Girvin, Steven M and Yoshioka, D},
  journal   = {Phys. Rev. B},
  volume    = {37},
  number    = {16},
  pages     = {9753},
  year      = {1988},
  doi       = {10.1103/PhysRevB.37.9753},
  publisher = {APS}
}

@article{bravyi2011schrieffer,
  title     = {{Schrieffer--Wolff} transformation for quantum many-body systems},
  author    = {Bravyi, Sergey and DiVincenzo, David P and Loss, Daniel},
  journal   = {Ann. Phys.},
  volume    = {326},
  number    = {10},
  pages     = {2793--2826},
  year      = {2011},
  doi       = {10.1016/j.aop.2011.06.004},
  publisher = {Elsevier}
}

@article{maruyama2026graph,
  title     = {Graph minor embedding can affect sampling degenerate ground states using quantum annealing},
  author    = {Maruyama, Naoki and Ohzeki, Masayuki and Tanaka, Kazuyuki},
  journal   = {J. Phys. Soc. Jpn.},
  volume    = {95},
  number    = {1},
  pages     = {013002},
  year      = {2026},
  doi       = {10.7566/JPSJ.95.013002},
  publisher = {The Physical Society of Japan}
}

@article{maruyama2026interpretation,
  title   = {Interpretation of Unfair Sampling in Quantum Annealing by Node Centrality},
  author  = {Maruyama, Naoki and Ohzeki, Masayuki},
  journal = {arXiv:2601.01920},
  year    = {2026}
}

@article{moessner2000two,
  title     = {Two-dimensional periodic frustrated {Ising} models in a transverse field},
  author    = {Moessner, R and Sondhi, Shivaji Lal and Chandra, P},
  journal   = {Phys. Rev. Lett.},
  volume    = {84},
  number    = {19},
  pages     = {4457},
  year      = {2000},
  doi       = {10.1103/PhysRevLett.84.4457},
  publisher = {APS}
}

@incollection{tanaka2000quantum,
  author    = {Tanaka, Shu and Tamura, Ryo},
  title     = {Quantum annealing: From viewpoints of statistical physics, condensed matter physics, and computational physics},
  booktitle = {Lectures on Quantum Computing, Thermodynamics and Statistical Physics},
  pages     = {1--59},
  doi       = {10.1142/9789814425193_0001},
  year      = {2012},
  publisher = {World Scientific}
}

@article{moessner2001ising,
  author  = {Moessner, R. and Sondhi, S. L.},
  title   = {Ising models of quantum frustration},
  journal = {Phys. Rev. B},
  volume  = {63},
  pages   = {224401},
  year    = {2001},
  doi     = {10.1103/PhysRevB.63.224401}
}

@article{hen2016driver,
  title     = {Driver {Hamiltonians} for constrained optimization in quantum annealing},
  author    = {Hen, Itay and Sarandy, Marcelo S},
  journal   = {Phys. Rev. A},
  volume    = {93},
  number    = {6},
  pages     = {062312},
  year      = {2016},
  publisher = {APS},
  doi       = {10.1103/PhysRevA.93.062312}
}

@article{king2016degeneracy,
  title     = {Degeneracy, degree, and heavy tails in quantum annealing},
  author    = {King, Andrew D. and Hoskinson, Emile and Lanting, Trevor and Andriyash, Evgeny and Amin, Mohammad H.},
  journal   = {Phys. Rev. A},
  volume    = {93},
  issue     = {5},
  pages     = {052320},
  numpages  = {12},
  year      = {2016},
  month     = {May},
  publisher = {American Physical Society},
  doi       = {10.1103/PhysRevA.93.052320},
  url       = {https://link.aps.org/doi/10.1103/PhysRevA.93.052320}
}

@article{jordan2008perturbative,
  title     = {Perturbative gadgets at arbitrary orders},
  author    = {Jordan, Stephen P. and Farhi, Edward},
  journal   = {Phys. Rev. A},
  volume    = {77},
  number    = {6},
  pages     = {062329},
  year      = {2008},
  publisher = {APS},
  doi       = {10.1103/PhysRevA.77.062329}
}

@article{amin2009first,
  title     = {First-order quantum phase transition in adiabatic quantum computation},
  author    = {Amin, M. H. S. and Choi, V.},
  journal   = {Phys. Rev. A},
  volume    = {80},
  number    = {6},
  pages     = {062326},
  year      = {2009},
  publisher = {APS},
  doi       = {10.1103/PhysRevA.80.062326}
}

@article{boixo2013experimental,
  title     = {Experimental signature of programmable quantum annealing},
  author    = {Boixo, Sergio and Albash, Tameem and Spedalieri, Federico M and Chancellor, Nicholas and Lidar, Daniel A},
  journal   = {Nat. Commun.},
  volume    = {4},
  number    = {1},
  pages     = {2067},
  year      = {2013},
  publisher = {Nature Publishing Group UK London},
  doi       = {10.1038/ncomms3067}
}

@article{kudo2018constrained,
  title     = {Constrained quantum annealing of graph coloring},
  author    = {Kudo, Kazue},
  journal   = {Phys. Rev. A},
  volume    = {98},
  number    = {2},
  pages     = {022301},
  year      = {2018},
  publisher = {APS},
  doi       = {10.1103/PhysRevA.98.022301}
}

@article{erdos1959random,
  title   = {On random graphs {I}},
  author  = {Erd{\H{o}}s, P. and R{\'e}nyi, A.},
  journal = {Publ. Math. Debrecen},
  volume  = {6},
  number  = {3--4},
  pages   = {290--297},
  year    = {1959},
  doi     = {10.5486/PMD.1959.6.3-4.12}
}

@article{gilbert1959random,
  author    = {Gilbert, Edgar N.},
  journal   = {Ann. Math. Stat.},
  number    = {4},
  pages     = {1141--1144},
  publisher = {Institute of Mathematical Statistics},
  title     = {Random Graphs},
  volume    = {30},
  year      = {1959},
  doi       = {10.1214/aoms/1177706098}
}

@article{tanaka2010nonmonotonic,
  title     = {Nonmonotonic dynamics in a frustrated {Ising} model with time-dependent transverse field},
  author    = {Tanaka, Shu and Miyashita, Seiji},
  journal   = {Phys. Rev. E},
  volume    = {81},
  number    = {5},
  pages     = {051138},
  year      = {2010},
  publisher = {APS},
  doi       = {10.1103/PhysRevE.81.051138}
}

\end{document}